\documentclass[onecolumn,superscriptaddress,notitlepage,nofootinbib,citeautoscript, showpacs,showkeys]{revtex4-1}

\usepackage[utf8]{inputenc}

\date{\today}
\usepackage{amsmath,amssymb}
\usepackage{epsfig}
\usepackage{epstopdf}
\usepackage{subfig}
\usepackage{graphicx}
\usepackage{dcolumn}
\usepackage{bm}
\usepackage[colorlinks,linkcolor=magenta,citecolor=magenta,hyperindex,CJKbookmarks]{hyperref}
\usepackage{float}
\usepackage{comment}
\usepackage{xcolor}
\usepackage{tabularx} 

\newcommand{\bst} {Ba$_2$ScTaO$_6$}
\newcommand{\gen} {A$_2$BB'O$_6$}

\makeatletter

\DeclareMathOperator{\sech}{sech}

\begin{document}

\title{Multiple low-energy excitons and optical response of \texorpdfstring{{$d^0$}}{d0} double perovskite \texorpdfstring{{\bst}}{bst}}

\author{A. K. Himanshu}
\email{akhimanshu@gmail.com}
\affiliation{Variable Energy Cyclotron Center (VECC), DAE, 1/AF Bidhannagar, Kolkata, India  700064}, 
\affiliation{Homi Bhabha National Institute, Mumbai, India 400094} 
 
\author{Sujit Kumar}
\affiliation{Department of Physics, Magadh University, Bodh Gaya,India 824234}
 
\author{Urmimala Dey}
\affiliation{Department of Physics, Indian Institute of Technology Kharagpur, Kharagpur ,India 721302}, 
\affiliation{Department of Physics, Durham University, South Road, Durham DH1 3LE, UK} 
 
\author{Rajyavardhan Ray}
\email{r.ray@bitmesra.ac.in}
\affiliation{Leibniz IFW Dresden, Helmholtzstr. 20, 01069 Dresden, Germany}
\affiliation{Dresden Center for Computational Materials Science (DCMS), TU Dresden, D-01062, Germany}
\affiliation{Department of Physics, Birla Institute of Technology Mesra,
Ranchi, India 835215}

\date{\today}

\begin{abstract}

Large bandgap insulators are considered promising for applications such
as photocatalysts, dielectric resonators and interference filters.
Based on synchrotron X-ray diffraction, diffuse reflectance
measurement and density functional theory, we report the crystal
structure, optical response, and electronic properties of the
synthesized $d^0$  double perovskite {\bst}. In contrast to
earlier prediction, the electronic bandgap is found to be large,
$\sim 4.66$ eV. The optical response is characterized by the presence of
multiple exciton modes extending up to the visible range. A
detailed investigation of the direct gap excitons based on the
Elliot formula is presented. Density functional theory based
investigation of the electronic properties within generalized
gradient approximation severely underestimates the electronic gap.
To reach a quantitative agreement, we consider different available
flavors of the modified-Becke-Johnson exchange-correlation
potential and discuss their effects on the electronic and optical
properties.

\end{abstract}


\maketitle

\section{Introduction} 

Double Perovskite oxides (DPOs), with the general formula {\gen},
constitute an important materials class exhibiting a wide variety of
interesting physical phenomena, like high temperature superconductivity,
colossal magnetoresistance, photocatalytic activity, high dielectric
constants, etc. \cite{Sleight1961,Patterson1963, Vasala2015,
SahaDasgupta2020, Eng2003, Ray2017, Yin2019}. In ordered systems, the
transition-metals B and B$’$ are surrounded by oxygen atoms in an
octahedral environment with B$-$O$-$B$'$ links. Therefore, DPOs consist
of alternating corner sharing BO$_6$ and B$'$O$_6$ octahedra stacked in
all three directions. On the other hand, the A-sites, centered at the
interstitial voids created by these octahedra, are typically occupied by
divalent alkaline earth metals or trivalent rare-earth metals. 
Depending on relative size of the cations to anions, different 
structures may be realized: ideally, DPOs have a cubic structure and is 
realized in most cases at high temperatures. However,
depending on the size, A-site cations may induce tilting and rotations
of the BO$_6$ and B$'$O$_6$ octahedra, leading to deviations from the
ideal cubic structure. The resulting low-symmetry structure could be
tetragonal, monoclinic, orthorhombic, or rhombohedral. A
schematic representation of an ordered cubic DPO is shown in Fig.
\ref{fig:str}(b). The presence of different metal sites (A-, B- and
B$'$-sites) allows a high degree of flexibility in crystal structure,
opto-electronic and magnetic properties, responsible for the mentioned
range of interesting phenomena. This further makes them appealing for
novel device applications.

The large-gap $d^0$-DPOs formed by transition metals with unfilled/empty
valence $d$-shells are of special interest \cite{Eng2003, Ray2017,
Yin2019}. They are considered promising in microwave-dielectric
resonator applications, including interference filters, reflective
coating and in optical fibers, due to their good dielectric properties,
and as buffer materials due to their low reactivity. The properties of
such materials are largely governed by the transition metal species and
structural distortions \cite{Eng2003, Ray2017}. For example, it was
found that the bandgap increases with increasing octahedral tilting and
distortions. Moreover, difference in the electronegativity of the B and B$'$ cations
can significantly affect the electronic bandgaps: compositional
modulations can alter the bandgap by up to 2 eV \cite{Eng2003}. Therefore, to ascertain
the feasibility of these materials for possible technological
applications, a detailed characterization of their optical response over
a wide frequency range is indispensable.

In this combined experimental and theoretical study, we focus on {\bst}.
Among the Ta-based DPOs, $A_2$MTaO$_6$, M = Sc compounds are slightly
different from other members of the family in terms of their bandgap
\cite{Eng2003}. Further, the electronic bandgap of {\bst} was predicted
to be the smallest (3.35 eV) within the $A_2$ScTaO$_6$ compounds. With
decreasing size of $A$ cations, octahedral rotations and distortions
induce a larger bandgap. The formal valencies of Sc and Ta are,
respectively, $+3$ and $+5$, leading to the $d^0$ configuration for both
the involved transition metals. The ionic radii of the transition metal
cations are found to be 0.745 {\AA} and 0.64 {\AA} for Sc$^{3+}$ and
Ta$^{5+}$, respectively. Large difference in the ionic radii is expected
to lead to a long-range ordering of the BO$_6$ and B$'$O$_6$ octahedra in
the crystal structure. A good measure of the crystal structure symmetry
of perovskite oxides is the tolerance factor
\cite{Vasala2015,SahaDasgupta2020,Ray2017,Eng2003,Ray2021,Travis2016}. For {\bst}, the
tolerance factor is found to be $t=1.02$ \cite{Eng2003}, implying the
likelihood of a cubic symmetry since $t \approx 1$.

The primary objective of this study was to ascertain the origin of low
predicted bandgap of {\bst}: whether this is an intrinsic property of
the material or arises due to DFT approximations. To this end, we report
the crystal structure, obtained via synchrotron X-ray diffraction
(SXRD), and the optical gap, obtained via the diffuse reflectance
measurement in the UV-Vis range. We find that {\bst} crystallizes in an
ordered cubic structure and has a gap of $\sim4.7$ eV, similar to other
$A_2$MTaO$_6$ (M = Y, Al and Ga). For the synthesized DPO {\bst}, we
have performed a thorough investigation of the electronic and optical
properties within the framework of density functional theory (DFT).
Standard DFT using generalized gradient approximation (GGA) agrees with
the earlier reports. A quantitative agreement between theory and
experiments is, however, found only upon including the modified
Becke-Johnson (mBJ) exchange-correlation potential. This leads to
accurate predictions of the optical response, especially beyond the
experimental range. Inclusion of the mBJ
potential reduces the conduction bandwidth of the $t_{2g}$ states and
increases the bandgap. In particular, our comparative study of different parametrizations of the mBJ potential clarifies the most
suitable choice of mBJ potential for {\bst} and likely also for other
materials with similar bandgap and electronic properties.

\section{Experimental and computational details}
\label{sec:methods}

\subsection{Synthesis and characterization}
\label{sec:methods-synthesis}
 
\begin{figure}[ht!]
    \begin{center}
    \includegraphics[scale=1.000,angle=0]{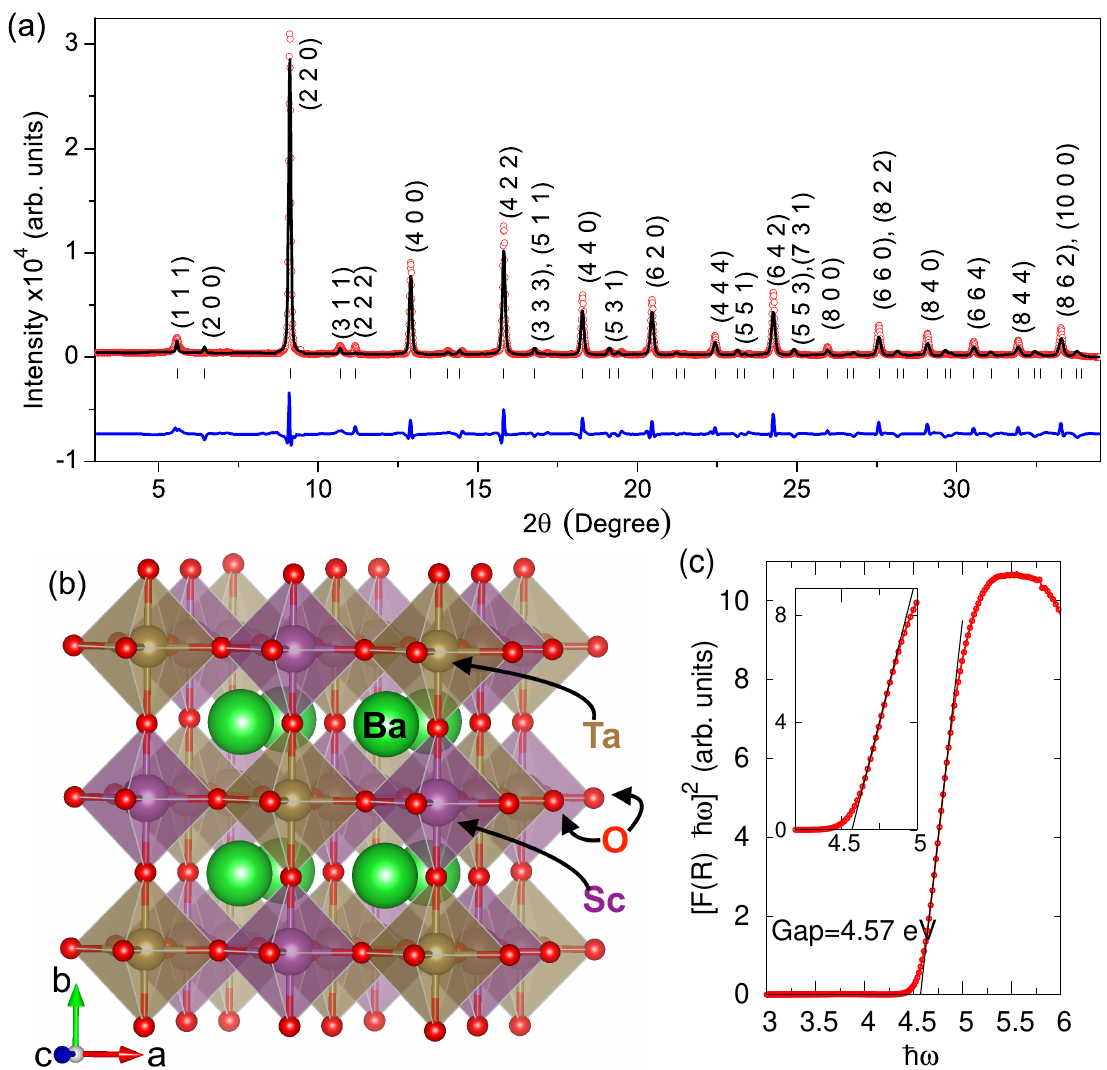}
    \end{center}
    \caption{(Color online.) (a) The SXRD pattern and (b) the crystal
      structure of {\bst}. The corresponding crystal structure
      parameters are provided in Table \ref{table:str}. (c) The
      dependence of the modified Kubelka-Munk (KM) function,
      corresponding to direct-allowed transitions, on the photon energy
      $\hbar\omega$ and the linear part showing an optical band-gap of 4.57 eV.
      The inset shows a zoom of the plot near the gap energy.}
    \label{fig:str}
\end{figure}

{\bst} was synthesized by conventional solid state 
reaction at 1500 $^{\circ}$C for 72 hours, from a mixture of BaCo$_3$ 
(99\%, Merc India), Sc$_2$O$_3$ (99.99\%, Alfa Aesar),
Ta$_2$O$_5$ (99.99\%, Aldrich) powders in stoichiometry proportions.
Color of the obtained
samples was found to be off-white. Synchrotron X-ray Diffraction (SXRD)
study was carried out using BL-11 of Indus-2 synchrotron source
\cite{BL11_Pandey2013}, using  $\lambda=0.46195$ {\AA} beam with a beam
current of 100 meV and energy 2.5 GeV, at the Raja Ramanna Center for
Advanced Technology (RRCAT), Indore, India. The Rietveld analysis
software FULLPROF \cite{fullprof} was used to analyze the SXRD data,
which shows a cubic structure, space group $Fm\bar{3}m$ (No. 225). Ba,
Sc, Ta, and O atoms occupy the Wyckoff positions of $8c$, $4a$, $4b$
and $24e$, respectively. The SXRD pattern and the Rietveld fit are shown
in Fig. \ref{fig:str}(a). The crystal structure is shown in Fig.
\ref{fig:str}(b) and the corresponding details related to the structural
and the refinement parameters are listed in Table \ref{table:str}.

\subsection{UV-Vis spectroscopy}
\label{sec:methods-optical}

Diffusive reflectance spectroscopy was carried out in the UV-Vis
range using the UV-Vis spectrophotometer Perkin-Elmer 950.
In order to estimate the electronic gap, the measured
reflectance spectrum was converted into an effective absorption spectrum
--- the Kubelka-Munk (KM)
function, $F(R)$ \cite{Davis1970, Barton1999}. Within this approach, the
electronic bandgap is
obtained using \cite{Ray2017}:
\begin{equation}
  F(R) \propto \frac{(\hbar\omega -E_g)^n}{\hbar\omega} \,, 
\label{eqn:km}
\end{equation}
with $n=1/2$ corresponds to direct-allowed transitions. $\hbar\omega$ and $E_g$
are the incident photon energy and the gap, respectively. As outlined in
\cite{Ray2017}, the intercept of the linear part of $[F(R) \hbar\omega]^2$ vs
$\hbar\omega$ curve on the energy axis is the bandgap. Figure \ref{fig:str}(c)
shows the absorption spectrum (KM function) and the
corresponding value of optical gap. The value of the bandgap was found to be 4.57 eV.

\begin{figure}[b!]
    \begin{center}
    \includegraphics[scale=0.450,angle=0]{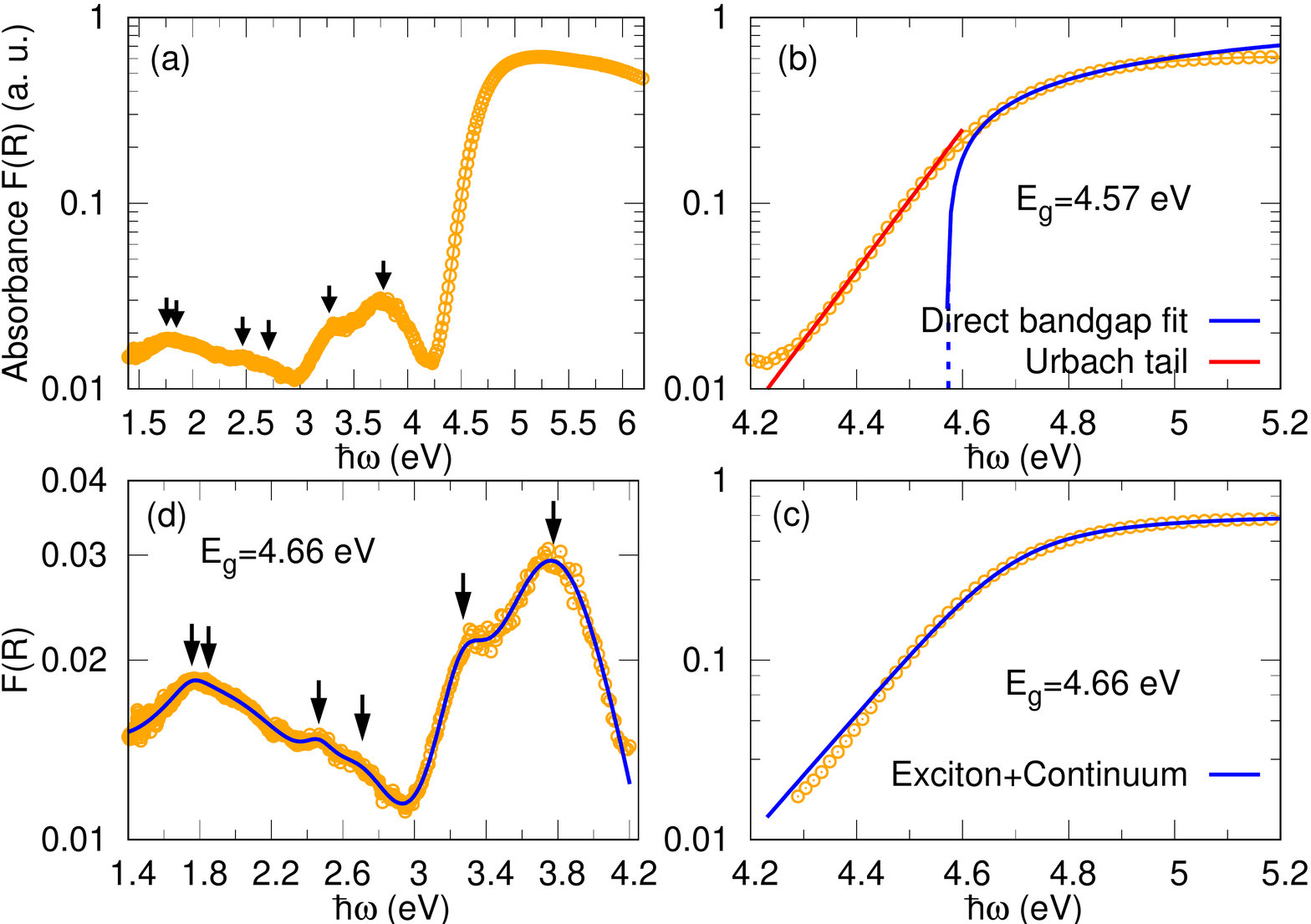}
    \end{center}
      \caption{(Color online.) (a) The absorption spectrum of {\bst}. The
      arrows indicate the exciton modes. (b)-(c) Fit of the high-energy part of the
      absorption spectrum to (b) direct-allowed optical transition with
      Urbach tail, (c) continuum model with and without exciton modes. (d) The low-energy part of the
      absorption spectrum (symbols) and a fit of the 6-peak exciton model
      (solid line).}
    \label{fig:optical}
\end{figure}

However, a closer look at the absorption spectrum (KM function, $F(R)$) in the
log-scale reveals multiple low-intensity peaks in the energy range
between 1.4 eV and 4.2 eV, shown in Fig. \ref{fig:optical}(a). These
peaks indicate presence of exciton modes. To obtain details regarding
these exciton modes, the 
absorption data was modeled based on the Elliot formula for direct-gap
excitons \cite{Elliot1957, Wang2017}. 
The absorption is thus given by a combination of excitonic
contributions, $\alpha_{nx}$, and a continuum, $\alpha_{\rm cont}$, 
contributing at low (band edge) and high energies, respectively \cite{Wang2017}:
\begin{align}
\thinmuskip=\muexpr\thinmuskip*4/8\relax
\medmuskip=\muexpr\medmuskip*4/8\relax
      \alpha(\omega) &= \sum_n \alpha_{nx}(\omega) + \alpha_{\rm cont}(\omega) \nonumber \\
       &= \sqrt{E_b}\Bigg[ \sum_n 2A_n\frac{E_b}{n^3}  \sech
       \bigg(\frac{ \hbar\omega - E_g + E_b/n^2}{\Gamma_n}\bigg) \nonumber \\
&+ B \int_{E_g}^{\infty} \sech \bigg(\frac{\hbar\omega -
      E}{\Gamma^{'}}\bigg)  \nonumber \\
      &\times \frac{1 + 10\frac{m^2}{\hbar^4}EC_{np} +
                     \big(\frac{\sqrt{126}m^2}{\hbar^4}EC_{np} \big)^2}{1 - \exp{\bigg( -2\pi
                     \sqrt{\frac{E_b}{E-E_g}}\bigg) } } dE \Bigg] \,.
\label{eqn:elliot}
\end{align}
Here, $n$ stands for the order of the exciton state. 
$\alpha_{nx}$, therefore, corresponds to the absorption from the $n$-th
exciton state. `$\sech$' function is used as a broadening function for the exciton
lineshape with a linewidth $\Gamma_n$. $E_g$ is the electronic bandgap
while $E_b$ is the exciton binding
energy. $C_{np}$ is the correction factor to account for deviations from
parabolic bands. $B$ and $\Gamma'$, respectively, denote the amplitude
and the broadening for the continuum part. $m$ is the free electron mass and $\hbar$ is the reduced Planck constant.

A least-squares fit of the above expression in Eq.
(\ref{eqn:elliot}) to the data was obtained using the python-lmfit
function and scipy (python version 3.8.12). For brevity, the low-energy part, up to 4.2 eV, consisting of multiple exciton peaks
(marked by arrows in Fig. \ref{fig:optical}(a)) and the high energy
part were considered separately. 
For the high energy part, 
above photon energy $\hbar\omega = 5.2$ eV, deviations
between the calculated and the observed spectrum were noted for all cases
discussed below. Therefore, only fitting up to 5.2 eV was considered.

Different models were considered for the high-energy part of the
absorption data: first, given the exponential onset of the absorption spectrum in this
energy window, thus describing the data by a Urbach tail [$U(\omega) =
U_0 exp\{(\omega - E_1)/E_{\rm U}\}$; $E_{\rm U}$ is the Urbach energy
while $E_1$ and $U_0$ are fitting parameters] and a direct
gap function [Eq. (\ref{eqn:km})], as shown in Fig.
\ref{fig:optical}(b). Second, a fit of $\alpha(\omega)$ consisting of one exciton mode (with a suitable value
of $n=7$) was also investigated. For the latter,
the best-fit was obtained for $E_g=4.66$ eV ($\chi^2 = 1.181 \times
10^{-3}$) and is shown in Fig. \ref{fig:optical}(c).

On the other hand, for the low-energy part of the data,  a fit of $\alpha^{\prime}(\omega)=\sum_n \alpha_{nx}$ was
successfully obtained with six exciton mode peaks. Furthermore, it should be noted that the data around the first peak, at
$\sim 1.76$ eV, is asymmetric. Therefore, to reliably obtain the value
of the peak position, an ``offset" to the sech function, leading to 
$\beta(\omega)=\sum_n \alpha_{nx} + k_n$, was also
considered. However, the resulting values of offsets were found to be
negative. Alternatively, we considered a hypothetical peak at
the edge of the data to account for the offset in only the first peak.
This immediately led to a remarkable fit to the data and is shown in
Fig. \ref{fig:optical}(d). The reliability of the estimation of the this
peak position was tested by considering an isolated fit of the
$\alpha^{\prime}(\omega)$ with a constant offset. 

The resulting best-fit ($\chi^2 = 7.41\times 10^{-5}$) corresponds
to $E_g = 4.66$ eV . However, the quality of the fit remained unchanged for small changes
in the values of $E_g$. For example, with $E_g=4.67, 4.68$ eV, the difference
in $\chi^2$ from the fit with $E_g=4.66$ is $\Delta
\chi^2 \sim 10^{-10}$.  

Other models, such as a five exciton peak model and four exciton peak models were also considered but led to a
poorer fit ($\chi^2_{\rm 5-peak} = 7.570\times 10^{-5}$; $\chi^2_{\rm 4-peak} = 8.196\times 10^{-5}$). In the following, we will,
therefore, consider the electronic gap of {\bst} to be $E_g = 4.66$
eV.

\subsection{Density functional theory}
\label{sec:methods-dft}
We adopted the all electron full-potential linearized augmented plane wave (FP-LAPW) method of DFT
with the scalar relativistic approximation, as implemented in the WIEN2k
code \cite{wien2k,Blaha2020}. All calculations were carried out for the
experimental crystal structure (Table \ref{table:str}),
using a $12\times12\times12$ $k$-mesh in the full Brillouin zone (BZ)
($\sim 72$ $k$-points in the irreducible Brillouin zone) to carry out
the integrals over the BZ. The muffin-tin radii for Ba, Sc, Ta, and O
were kept fixed at 2.5 a.u., 2.3 a.u., 2.03 a.u. and 1.54 a.u.,
respectively. The $R_{MT} \times k_{\rm max}$ parameter was chosen to be
7.0, where $k_{\rm max}$ is the plane-wave cut-off and $R_{MT}$ is the
smallest muffin-tin radii among all atoms. The exchange and correlation
effects have been treated within the Perdew-Burke-Erzenhof (PBE)
implementation of the Generalized Gradient Approximation (GGA)
\cite{Perdew1996}. Self-consistent solutions correspond to convergence
below 0.0001 e/a.u.$^3$ for charge, and 0.01 mRy for the total energy
per unit cell.

In order to address the well-known issue of bandgap underestimation
within DFT \cite{Baerends2013}, calculations were also performed using
the semi-local Tran-Blaha modified Becke-Johnson (TB-mBJ)
exchange-correlation potential \cite{Tran2009}, as implemented in
WIEN2k. Different implementations of the mBJ potential available in
WIEN2k were also tested: in particular, the refined mBJ potentials
suitable for perovskite oxides \cite{Koeller2012}, large bandgap solids
\cite{Koeller2012} and halide perovskites \cite{Jishi2014}.

The optical properties were obtained using complex dielectric function 
\begin{equation}
    \varepsilon(\omega) = \varepsilon_1(\omega) +i \varepsilon_2(\omega)\,.
\end{equation}
In general, $\varepsilon$ is a second rank tensor with nine independent components. However, depending on the crystalline symmetry, only a few of these components could be independent. For a cubic structure, for example, the three principal directions ($x$, $y$, and $z$) are equivalent, leading to only one independent value.
The real and imaginary part of the dielectric response function are related by Kramers-Kronig relation:
\begin{equation}
    \varepsilon_{1,ij} (\omega) = \delta_{ij} +  \frac{2}{\pi}\,\,\mathcal{P} \int_0^{\infty} \frac{\omega'\,\varepsilon_{2,ij}(\omega')}{\omega'^2 -\omega^2} d\omega' \,
\end{equation}
where, $\mathcal{P}$ stands for the principal part of the integral and $\varepsilon_{a,ij}$ is the $ij$-th component of $\varepsilon_a$ ($a=1,2$).
All other optical response functions, such as the refractive index, optical conductivity, absorption coefficient and the electron loss function, were obtained by using their well-defined mathematical relations with the dielectric function  \cite{Draxl2006}. The broadening parameter was chosen to be 0.1 eV.

\section{Results and discussions}

\subsection{Crystal structure}

\bst\ crystallizes in a cubic $Fm{\bar 3}m$ structure (No. 225)  with lattice constant $a=8.226$ {\AA}. The only free parameter in this structural model, {\it viz.} the $x$-coordinate for the O atom is found to be $x=0.245$ based on a high-resolution SXRD data. The complete details of the crystal structure are listed in Table \ref{table:str}. Within the cubic symmetry, the Sc-O-Ta angle is 180$^{\circ}$. The M-O bond lengths are 2.0894 {\AA} and 2.0236 {\AA}, for M= Sc and Ta, respectively. These values are in excellent agreement with the earlier reported X-ray diffraction data in the Springer Materials database \cite{SpringerMaterials,Filipev1966}, but deviate slightly from the prediction based on SPuDS \cite{Lufaso2006}. 

\begin{table}[ht]
    \centering
    \caption{Crystal structure and refinement parameters for {\bst} obtained from SXRD data at room
    temperature. The atomic positions are in fractions coordinates ($x$,$y$,$z$) along with their
Wyckoff positions (WPs). } 
    \label{table:str}
    \begin{tabularx}{0.75\textwidth}{p{1.1cm} p{1.0cm} p{1.25cm} p{0.9cm} p{0.9cm} p{1.5cm}}
        \hline\hline 
        \multicolumn{6}{l}{\textbf{A. Refinement parameters}} \\
        \hline\hline 
        \multicolumn{3}{l}{\textbf{Params.}} & \multicolumn{3}{l}{\textbf{Value}} \\
        \hline\hline 
        \multicolumn{3}{l}{Space group} & \multicolumn{3}{l}{$Fm{\bar 3}m$ (No. 225)} \\
        \multicolumn{3}{l}{$a=b=c$ ({\AA})} & \multicolumn{3}{l}{8.226} \\
        \multicolumn{3}{l}{$\alpha=\beta=\gamma$} & \multicolumn{3}{l}{90$^{\circ}$} \\
        \multicolumn{3}{l}{Vol ({\AA}$^3$)} & \multicolumn{3}{l}{556.5(4)} \\
        \multicolumn{3}{l}{Formula units/unit cell} & \multicolumn{3}{l}{$Z=4$} \\
        \multicolumn{6}{l}{Final $R$-indices:}  \\
        \multicolumn{3}{l}{$R_{\rm p}$} & \multicolumn{3}{l}{4.17\%} \\
        \multicolumn{3}{l}{$R_{\rm wp}$} & \multicolumn{3}{l}{3.79\%} \\
        \multicolumn{3}{l}{$R_{\rm exp}$} & \multicolumn{3}{l}{2.81\%} \\
        \multicolumn{3}{l}{$\chi^2$} & \multicolumn{3}{l}{2.35} \\
        \hline\hline 
        \multicolumn{6}{l}{\textbf{B. Structural parameters}} \\
        \hline\hline 
        \textbf{Atoms} & \textbf{WP} & $\mathbf{x}$ & $ \mathbf{y}$ & $\mathbf{z}$ & \bf{B} ({\AA}) \\
        \hline\hline 
         Ba & $8c$ & 1/4 & 1/4 & 1/4 & 0.12(2) \\
         Sc & $4b$ & 1/2 & 1/2 & 1/2 & 0.11(1) \\
         Ta & $4a$ & 0 & 0 & 0 & 0.23(1) \\
         O & $24e$ & 0.245(3) & 0 & 0 & 0.08(0) \\
         \hline
    \end{tabularx}
\end{table}

\subsection{Exciton modes, optical and electronic gaps}
\label{sec:excitons}

UV-Vis spectroscopy is a convenient way of obtaining the
electronic/optical gap of a semiconductor. The simplest way to estimate
the electronic bandgap is through the KM function [Eq. (\ref{eqn:km})].
Figure \ref{fig:str}(c) shows the modified KM function corresponding
to direct-allowed transitions (see Sec. \ref{sec:methods-optical}) as
a function of the incident photon energy. The bandgap is obtained by
taking the intercept of the low-energy linear part of this curve on the
energy axis \cite{Davis1970, Barton1999, Ray2017}, which is estimated to be 4.57 eV. 

In general, however, the electronic bandgap, defined as the energy difference
between the valence band maxima and the conduction band minima,
can turn out to be significantly different from the optical gap due to presence
of exciton modes, $d-d$ transitions, phonon absorption and emission,
etc. \cite{King2009, Chiodo2010}. In such cases, the above analysis breaks down
rendering the estimate of the electronic gap inaccurate. 

The absorption spectrum (KM function) of {\bst} in log-scale, shown in Fig.
\ref{fig:optical}(a), reveals that the low-energy part up to $\sim 4.2$
eV contains multiple low-intensity peaks corresponding to exciton modes, marked by
arrows. A high quality fit of $\alpha_{nx}(\omega)$ [Eq. (\ref{eqn:elliot})] 
can be obtained with six exciton peaks ($n = 1 \ldots 6$), at
approximately 1.76 eV, 1.85 eV, 2.47 eV, 2.71 eV, 3.27 eV and 3.78
eV. Correspondingly, the renormalized exciton binding energies, $E_b/n^2$,
are 2.904 eV, 2.81 eV, 2.20 eV, 1.95 eV,
1.39 eV and 0.88 eV, respectively. The electronic bandgap $E_g$ is
found to be 4.66 eV, and the resulting fit is shown in Fig. \ref{fig:optical}(d).

The high-energy part, corresponding to photon energies above
4.2 eV is characteristically defined by a continuum absorption spectrum.
To model this part of the spectrum, different models were considered: 
first, driven by the
exponential tail of the absorption data in this energy window, we
considered the simplistic view based on direct-allowed transitions
without excitons. The band edge part of the data is modeled as an
exponential tail, so called Urbach tail, usually ascribed to shallow
traps and disorder \cite{John1986Urbach,Viljanen2020}. The Urbach energy
was obtained to be 114.42 meV from our room temperature data. The
corresponding fit is shown in Fig. \ref{fig:optical}(b).  
While it is apparent that the data can be described reasonably well, the estimated gap is at odds with the
estimated (electronic) bandgap $E_g=4.66$ eV obtained from the
low-energy part. Moreover, due to the presence of the excitons, the
continuum absorption spectrum is also affected \cite{Elliot1957,Wang2017} which is not accounted
for in this model. 

Later, the full expression for $\alpha(\omega)$ given in Eq.
(\ref{eqn:elliot}) was considered with one (possible) exciton mode
($n=7$) and $E_g = 4.66$ eV. 
The best-fit, shown in Fig. \ref{fig:optical}(c), corresponds to $\frac{m^2}{\hbar^4}C_{np}$ = -0.003
eV$^{-1}$, showing that there is no significant deviation from parabolic
bands.
The exciton binding energy was found to be 0.11 eV.
Small deviations are, however, observed in the energy window between 4.2 eV and 4.5 eV.
We, therefore, conclude that the electronic gap of {\bst} is 4.66 eV while
the optical gap is smaller $\sim 1.4$ eV.

The presence of exciton modes in the visible range (1.75 eV to 3.1 eV)
has an important consequence: based on the estimated
gap of 4.66 eV, {\bst} samples should be white in color due to lack of
absorption in the visible range. However, even if of low intensity,
presence of the exciton modes in the green ($\sim 2.3$ eV) and red
($\sim 1.8$ eV) regions, implies small absorption leading to the
off-white color.

\subsection{Electronic structure}

\begin{figure*}[t!]
    \begin{center}
    \includegraphics[scale=1.25,angle=0]{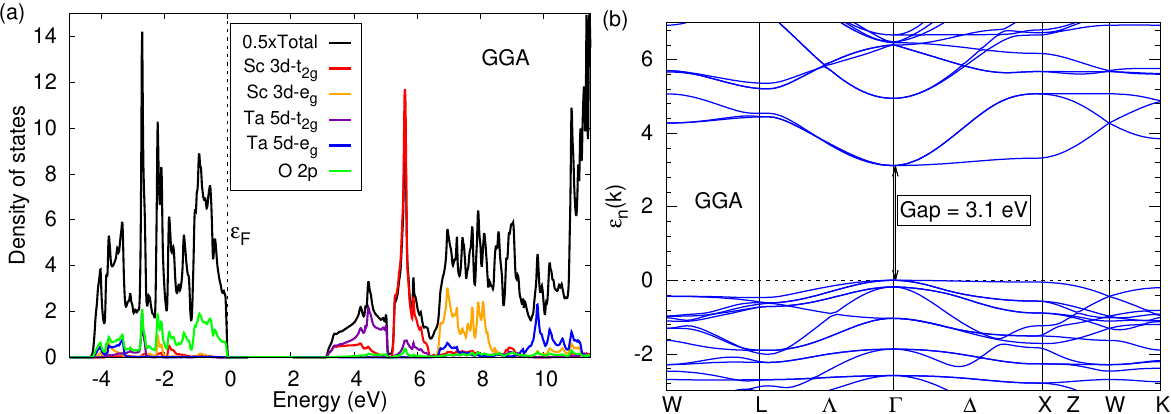}
    \caption{(Color online.) (a) The total and partial densities of states in states/eV-unit cell and states/eV-atom, respectively, obtained within GGA. The corresponding bandstructure showing the direct gap.}
    \label{fig:elprop-gga}
    \end{center}
\end{figure*}

DFT calculations using GGA lead to a non-magnetic insulating ground state for {\bst}.
The corresponding total and partial densities of states (DOS) are shown
in Figure \ref{fig:elprop-gga}(a), clearly exhibiting the insulating
nature.
The electronic properties of insulator DPOs can be well-described by the
crystal field experienced by the M-$d$ states and hybridization between
the oxygen $2p$ states and the M-$d$ states.
The crystal field produced by the nearest oxygen atoms in an octahedral
geometry lifts the five-fold degeneracy of M-$d$ states and leads to
splitting into the lower lying $t_{2g}$ states, consisting of $d_{xy}$,
$d_{yz}$, $d_{xz}$ orbitals, and $e_g$ states, consisting of
$d_{x^2-y^2}$ and $d_{z^2}$ orbitals. 
Typical of DPOs, hybridization with the surrounding oxygen atoms leads
to bonding states with dominant contribution from the O-$p$ orbitals in
valence band  while the conduction band edge comprises of  anti-bonding
states with dominant $d$ contribution. The relative energy position of
the transition metal $d$ states, however, depends on the metal species
(electronegativity, ionic radii, etc.) and occupation of the $d$ shell.

Indeed, for \bst, where both the transition metals are formally
unoccupied ($d^0$-ness), the M-$d$ states (both $t_{2g}$ and $e_g$
states) for M = Sc as well as M = Ta lie in the conduction band, as
shown in Fig. \ref{fig:elprop-gga}(a). This, in turn, confirms the
formal valencies of $+3$ and $+5$ for Sc and Ta, respectively.

\begin{figure*}[t!]
    \begin{center}
   \includegraphics[scale=1.25,angle=0]{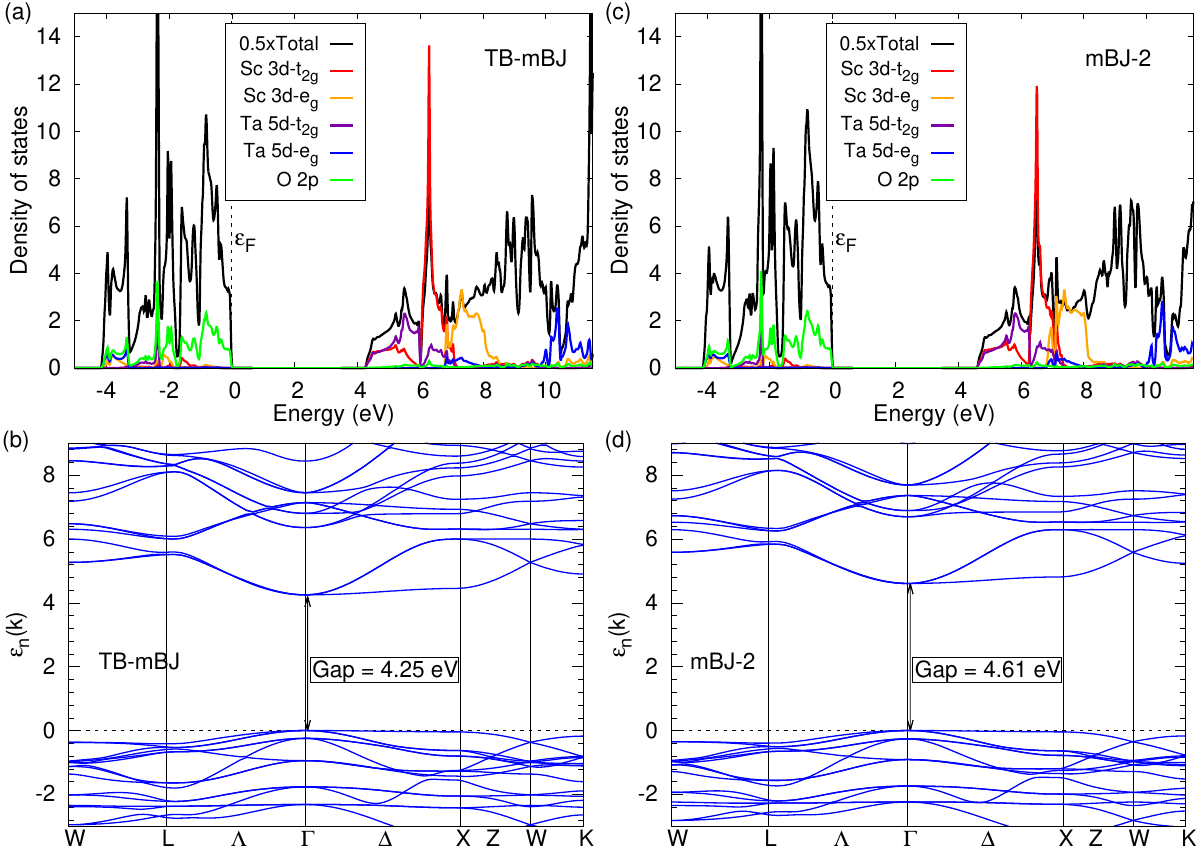}
    \caption{(Color online.) The total and partial density of states in states/eV-unit cell and states/eV-atom, respectively, and the corresponding bandstructure obtained within (a)-(b) TB-mBJ and (c)-(d) mBJ-2 method.}
    \label{fig:elprop-mbj}
    \end{center}
\end{figure*}
As observed from the atom- and orbital-resolved DOS,
the conduction band edge states have dominant contributions from the
Ta-$t_{2g}$ states with small contributions from Sc-$t_{2g}$ states.
This is somewhat different from $A_2$YTaO$_6$ DPOs \cite{Eng2003}
despite similar electronegativity of Y (1.22) and Sc (1.36). In
comparison, the electronegativity of Ta is 1.50. The primary reason for
this difference is the relatively small energetic overlap between the
Y-$4d$ and O-$2p$ states as compared to Sc-$3d$ and O-$2p$ states.
Strong covalency effects between Sc-O and Ta-O are also noted.
The dominant Ta-$t_{2g}$ and Ta-$e_{g}$ contributions are found at $\sim
4.5$ eV and $\sim 10$ eV, respectively, implying a crystal field
splitting of approximately 5.5 eV. 
On the other hand, the crystal field splitting for Sc-$d$ states is
merely $\sim 2.2$ eV --- a consequence of relative electronegativity
difference and M-O distances. The relative hybridization between Sc-O is
found to be quite smaller than Ta-O as evidenced by the M-$d$ DOS in the
valence band region. Therefore, an appropriate description of the
electronic structure is (large) mixing of Sc-$3d$ states  with the
strongly hybridized Ta-O anti-bonding states.

As shown in Fig. \ref{fig:elprop-gga}(b), the electronic bandstructure
within GGA has a direct gap of $\sim 3.1$ eV, which is much smaller than
the gap estimated from the analysis of the absorption data obtain
through UV-Vis reflectance spectroscopy.
This discrepancy between the theoretical and experimental estimates is a
well-known issue within DFT.
A computationally efficient way to address this discrepancy is to employ
the Tran-Blaha modified-Becke-Johnson (TB-mBJ) exchange-correlation
potential \cite{Tran2009}.  

Inclusion of such correction leads to a remarkable improvement in the
comparison between the measured and the calculated electronic gap, as
shown in Fig. \ref{fig:elprop-mbj}(a) \& (b). The electronic bandgap
within TB-mBJ is approximately 4.25 eV, much closer to the experimental
value (only $\sim 7\%$ smaller). Other aspects of the electronic
properties only undergo quantitative changes. The conduction band edge
has relatively large contribution from Sc-$d$ orbitals, implying further
enhancement of covalency effects. This is accompanied by reduction of
the effective crystal field splitting for the Sc-$d$ states to $\gtrsim
1$ eV. For Ta-$d$ states in comparison, the crystal field splitting
undergoes only marginal change. The bandwidth of the $d$ states, for
both Sc and Ta, is slightly reduced. It is important to note that these
corrections are not a rigid shift of all the conduction band states
("scissors" operation), and, therefore, may significantly influence the
optical properties. In comparison to GGA, a simplified view of the
application of the orbital-dependent corrections introduced by the mBJ
potential could be a band-dependent scissors operation like shifts
\cite{Sadhukhan2020}.

Despite its successes, TB-mBJ potential is known to underestimate the
gap in many cases \cite{Koeller2012}.
Different parametrizations of the mBJ potential have, therefore, been
suggested which are specifically catered to perovskite oxides
\cite{Koeller2012}, halide perovskites \cite{Jishi2014}, and large
bandgap materials \cite{Koeller2012}.
To further improve the degree of comparison between calculated and
measured gap, we first tried the mBJ potential adapted for perovskite
oxides (labeled mBJ-1). The resulting bandgap is found to be
approximately 4.36 eV. While the situation is improved as compared to
TB-mBJ, surprisingly it is still $\sim 4.5\%$ smaller. On the other
hand, the version adapted for perovskite halides (labeled mBJ-3), leads
to an overestimation presumably due to large electronegativity
differences between ligand species (oxygen vs halides). The resulting
bandgap is found to be $\sim 5.5$ eV. An excellent agreement is,
however, found for the mBJ potential adapted for large bandgap
insulators (labeled mBJ-2). These results are summarized in Table
\ref{table:mbj} for brevity, with details related to their
implementation in WIEN2k.

\begin{table*}[t!]
\renewcommand\arraystretch{1.25}
    \centering
    \caption{Electronic bandgap obtained within different DFT approximations.}
    \label{table:mbj}
    \begin{tabularx}{1.00\textwidth}{p{2.5cm} p{2.0cm} p{8.00cm} }
        \hline\hline 
          \textbf{Method} & \textbf{Gap} (eV) & \textbf{Remarks}  \\
        \hline\hline 
        Experimental & 4.66 & UV-Vis spectroscopy \\
        PBE-GGA & 3.12 &  \\
        TB-mBJ  & 4.25 & ``Option 0" in WIEN2k, defined in Ref. \cite{Tran2009}.  \\
        mBJ-1  & 4.36 & modified mBJ parameters suitable for perovskite oxides; \newline ``Option 1" in WIEN2k, defined in Ref. \cite{Koeller2012}. \\
        mBJ-2 & 4.61 & modified mBJ parameters suitable for large bandgap materials; \newline ``Option 2" in WIEN2k, defined in Ref. \cite{Koeller2012}. \\
        mBJ-3 & 5.48 & modified mBJ parameters suitable for halide perovskites \newline ``Option 4" in WIEN2k, defined in Ref. \cite{Jishi2014}. \\
         \hline
    \end{tabularx}
\end{table*}

The bandgap within mBJ-2 is found to be 4.61 eV which is in
excellent agreement with the experimental estimate of 4.66 eV. In
terms of the electronic properties, as shown in Fig.
\ref{fig:elprop-mbj}(c) and (d), DOS and bandstructures are
qualitatively similar to TB-mBJ and shows only small quantitative
changes. In view of this, to study the optical
properties, we consider only the mBJ-2 case.

\subsection{Optical properties}

\begin{figure*}[t!]
    \centering
    \includegraphics[scale=0.5,angle=0]{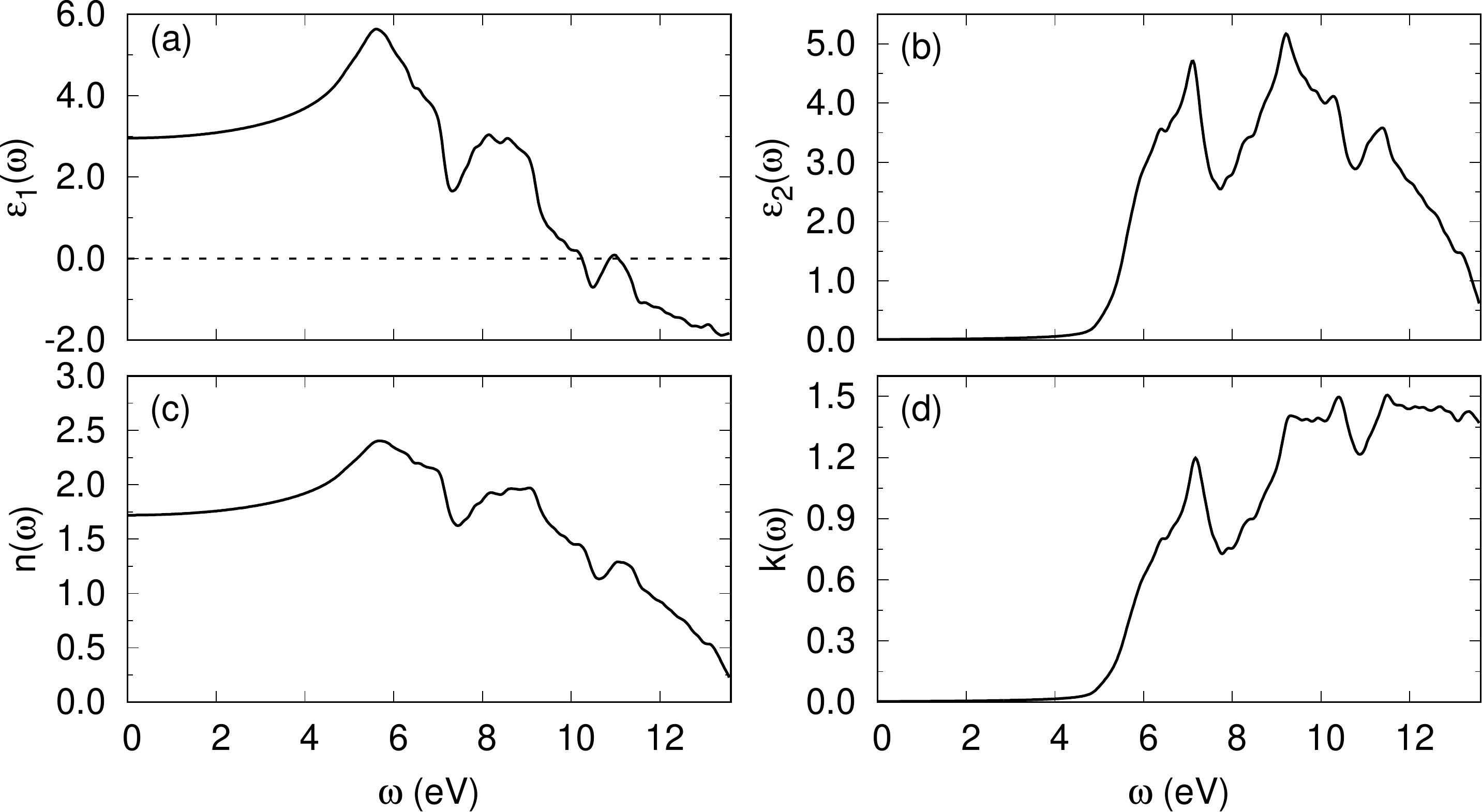}
    \caption{The (a) real [$\varepsilon_1(\omega)$] and the (b) imaginary [$\varepsilon_2(\omega)$] parts of the dielectric function. The (c) real [$n(\omega)$] and the (d) imaginary [$k(\omega)$] parts of the refractive index.}
    \label{fig:epsilon}
\end{figure*}
\begin{figure*}[t!]
    \centering
    \includegraphics[scale=0.5,angle=0]{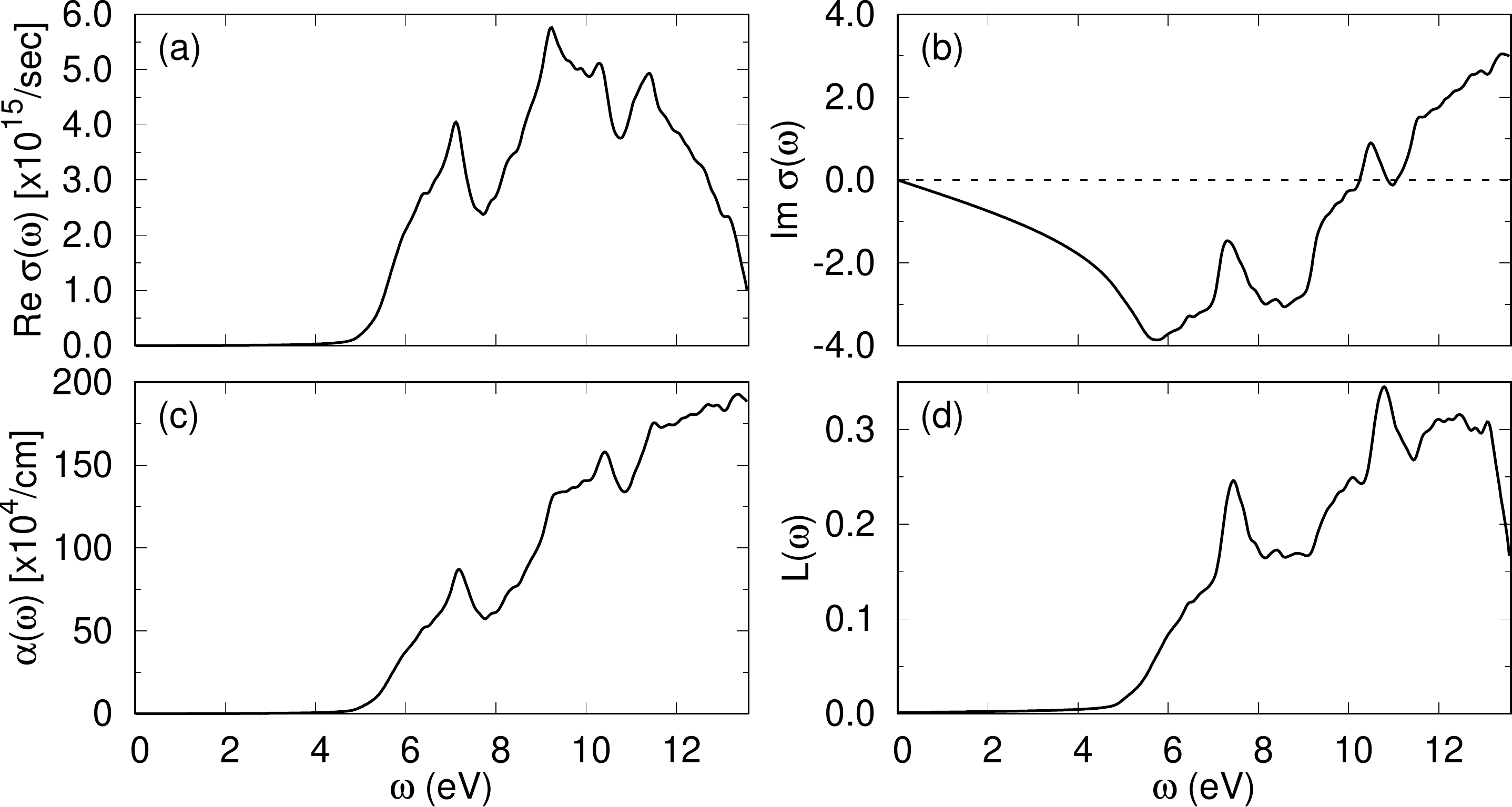}
    \caption{The (a) real and the (b) imaginary parts of the optical conductivity $\sigma (\omega)$, and the (c) absorption coefficient, and the (d) loss function.}
    \label{fig:cond}
\end{figure*}

Finite-frequency optical response of materials is captured through their
complex dielectric function $\varepsilon_{\mu \nu}(\omega)$ ($\mu,\nu =
x,y,z$). For structures with cubic symmetry, only one component of
$\varepsilon_{\mu \nu}$ is independent as $\varepsilon_{xx} =
\varepsilon_{yy} = \varepsilon_{zz}$. The symmetry relation between
different components also holds for other optical response functions.

In the case of {\bst}, the band edge region consists of multiple
excitons (see Sec. \ref{sec:excitons}) which have not been taken into
account in the DFT calculations.
However, away for the band edge, accurate predictions can be obtained
using the mBJ-2 corrections to DFT.

The dielectric function for {\bst} for the mBJ-2 method 
is shown in Fig. \ref{fig:epsilon}(a)-(b).
The onset of the optical response $\varepsilon_2$ is at a finite
frequency, at $\sim 4.6$ eV, due to the finite electronic bandgap. The
peaks in $\varepsilon_2(\omega)$ correspond to optical transitions from
states in the valence band to states in the conduction band and are
proportional to the joint DOS between the initial and final states
involved in such transitions.
In general, the optical selection rules dictate if such transitions are
allowed. In the present case, as also for most insulating DPOs, the
valence band and conduction band states close to the Fermi energy are
dominantly of O-$2p$ and M-$d$ characters. As a result, the selection
rules are trivially satisfied. 
Indeed, the transitions between the O-$2p$ states in the valence band
and M-$d$ states at the conduction band edge lead to the optical
transitions in the entire energy range studied.
For example, the features below $\sim 6.5$ eV can be understood as
transitions from the valence band region to the conduction band edge
(Ta-$t_{2g}$ states). Similarly, prominent peaks in $\varepsilon_2$ at
$\sim 7$ eV and $\sim 11.5$ eV arise due to transitions from valence
band edge to the Sc-$t_{2g}$ states and transitions from valence band
edge to Ta-$e_{g}$ states, respectively.

The frequency-dependence of the real and imaginary parts of the
refractive index closely follows the complex dielectric function, as
shown in Fig. \ref{fig:epsilon}(c)-(d) . They, respectively, correspond
to the dispersion and absorption of light passing through the medium.
The refraction coefficient peaks at the photon energy of 5.5 eV
approximately and reaches a maximum value of $\sim 2.4$. 
The static refractive index, $n(\omega = 0)$, is obtained from the
static limit of the real part of the dielectric function: $n =
\sqrt{\varepsilon_1(\omega \rightarrow 0)}$, and is found to be 1.72.

The complex optical conductivity $\sigma(\omega) = i \omega
\varepsilon(\omega)/4\pi$ is shown in  Fig. \ref{fig:cond}(a)-(b). The
low-energy peaks in the real part of the optical conductivity exhibit
characteristics similar to the dielectric function
$\varepsilon_2(\omega)$, with the prominent peaks at $\sim 7$ eV, $\sim
9.3$ eV, $\sim 10$ eV and $\sim 11.5$ eV. The dispersive part of the
dielectric function, $\varepsilon_1(\omega)$ governs the imaginary part
of the optical conductivity.
It is negative for small values of photon energy and crosses zero around
10 eV. The zero-crossing energy is consistent with the that of
$\varepsilon_1(\omega)$, as expected.

As shown in Fig. \ref{fig:cond}(c), the absorption edge lies at $\sim
4.6$ eV and increases with increasing photon energy. The finite value of
the absorption edge is due to the bandgap. The peaks around the energies
7 eV and above arise due to inter-band transitions, as discussed
earlier.

The Loss function, $L(\omega)$, corresponds to the electron energy loss
spectroscopy (EELS) which captures both non-scattering and elastic
scattering processes. 
It is connected to the energy loss of the electrons as they traverse
through the medium and lose energy. Typically, the mixing of single
electron excitations with the collective excitations (plasmons) gives
rise to energy loss of the electrons up to 50 eV.

The loss function for {\bst} is shown in Fig. \ref{fig:cond}(d) and
qualitatively follows the imaginary part of the dielectric function
$\varepsilon_2(\omega)$. There are two prominent peaks, lying at $\sim
7$ eV and $\sim 10.6$ eV. The first (low-energy) peak at $\sim 7$ eV,
originates from inter-band transitions between O-$2p$ to M$-d$ states as
discussed earlier, while the latter originates from transitions between
semi-core states to conduction band edge.

\section{Conclusions}
To summarize, we have experimentally determined the crystal
structure, optical and electronic gap of {\bst} and theoretically
investigated its electronic and optical properties. 
Contrary to the earlier prediction based on GGA, the experimental
electronic gap of {\bst} is found to be $\sim 4.7$ eV, which is similar to
other Ta-based $d^0$-DPOs. An excellent quantitative agreement between
experiment and DFT is obtained upon considering also the adapted mBJ
potential suggested for large gap insulators. Other mBJ potential
considered in this study, although better than GGA, still performs
rather poorly, including the one adapted for perovskite oxides.

An important finding of this study is the presence of multiple
low-energy exciton modes in {\bst} which extend well into the visible
range of the electromagnetic spectrum, thus affecting its optical
response. This may have remarkable consequences for other predicted
large bandgap DPOs, especially Ta-based DPOs.
Such materials are considered potential functional materials and may find
use as microwave dielectric resonators, interference filters, buffer
materials as well as photocatalysts.
However, to ascertain their utility, the electronic and optical properties should
perhaps be revisited in the light of our findings.

\subsection*{Acknowledgements}

AKH and SK thank Dr. V. Srihari for the SXRD experimental support at
BL-11, RRCAT, Indore. RR and UD thank Ulrike Nitzsche for technical assistance
with the computational resources in IFW Dresden. RR also thanks Dhiraj
Kushvaha for technical assistance.

\subsection*{Competing Interests} The authors declare that they have no known competing financial interests or personal relationships that could have appeared to influence the work reported in this paper.

\subsection*{Author contributions} AKH and SK carried out the synthesis
and characterization of the material. AKH and RR carried out the DFT
calculations. UD carried out the exciton analysis with help from RR. UD and RR
prepared the manuscript with contributions from all authors. AKH and RR
were responsible for project planning.

\subsection*{Data availability} The raw/processed data required to reproduce these findings cannot be shared at this time as the data also forms part of an ongoing study.








\end{document}